\documentclass[twocolumn,showpacs,preprintnumbers,amsmath,amssymb]{revtex4}

\usepackage{graphicx}
\usepackage{dcolumn}
\usepackage{bm}

\begin{document}


\title{Geometric phases and hidden gauge symmetry}

\author{Kazuo Fujikawa}
\affiliation{%
Institute of Quantum Science, College of Science and Technology,
Nihon University, Chiyoda-ku, Tokyo 101-8308, Japan
}%


\begin{abstract}
 The second quantized approach to geometric phases is reviewed.
The second quantization generally induces a hidden local (time-dependent) gauge symmetry. This gauge symmetry defines the  parallel transport and holonomy, and thus it controls
all the known geometric phases, either adiabatic or non-adiabatic, in a unified 
manner. The transitional region from the adiabatic to non-adiabatic phases is thus analyzed in a quantitative way. It is then shown that the topology of the adiabatic Berry's phase is trivial in a precise sense and also the adiabatic phase is rather fragile against the non-adiabatic deformation. In this formulation, the notion such as 
the projective Hilbert space does not appear.
\end{abstract}

\maketitle

\maketitle

\section{Introduction}

The  geometric phases\cite{1975Higgins,1984Berry,1987Berry}, which are often called Berry's phases, received much attention recently\cite{1992Anandan}. The purpose of the present article is to review a unified treatment of all the geometric phases, either adiabatic \cite{1975Higgins,1984Berry,1987Berry} or non-adiabatic\cite{1987Aharonov, 1988Samuel}, on the basis of the second quantized formulation\cite{2005Deguchi, 2005Fujikawa, 2007Fujikawa}. A characteristic feature of the geometric phases is that a certain kind of time-dependent symmetry is associated with them. This time dependent arbitrary freedom is important to define physical geometric phases, and it is generally referred to as  gauge symmetry. The local gauge symmetry in the conventional sense means the space-time dependent phase and thus the time dependent phase freedom associated with geometric phases is not actually local gauge symmetry, but following the convention in this field we simply call it local gauge symmetry.
The main purpose of the present article is to show that the origin of this gauge symmetry arises from the second quantization. Namely, the second quantization itself induces certain gauge symmetry. We use only the very
elementary aspect of the second quantization to show that this {\em hidden gauge symmetry} is crucial to define parallel transport and holonomy and thus it characterizes all the geometric phases.

We here recall that the geometric phases, either adiabatic or non-adiabatic, are associated with the time development of the state vector typically during one cyclic evolution. The time development of the state vector is described by the Schr\"{o}dinger equation, and the time development is entirely generated by the Hamiltonian. The evolution operator, which is generically defined by $|\psi(t)\rangle=U(t,0)|\psi(0)\rangle$, if evaluated exactly, thus contains all the information about the geometric phases. The primary object of our study is thus the evolution operator.

\section{Second quantization and hidden gauge symmetry}

In the second quantization we study the action
\begin{eqnarray}\label{eq-action1}
S&=&\int dtd^{3}x\left[\hat{\psi}^\dag(t.\vec{x})\Big(i\hbar\frac{\partial}{\partial
t}-\hat{H}(t)\Big)\hat{\psi}(t,\vec{x})\right]
\end{eqnarray}
for a time-dependent Hamiltonian $\hat{H}(t)$. We then expand
\begin{eqnarray}
\hat{\psi}(t,\vec{x})=\sum_{n}\hat{c}_{n}(t)v_{n}(t,\vec{x})
\end{eqnarray}
in terms of a suitable complete set $\{v_{n}(t,\vec{x})\}$ with
\begin{eqnarray}
 \hspace{1cm} \int d^{3}xv_{n}^{\star}(t,\vec{x})v_{m}(t,\vec{x})=\delta_{n,m}.
\end{eqnarray} 
A salient feature of the present formulation is that 
we allow the time-dependent complete set in the above expansion.
For the fermion, one imposes the equal-time 
anti-commutation relation
\begin{eqnarray}
\big\{\hat{c}_l(t),\hat{c}_m^\dag(t)\big\}=\delta_{lm}
\end{eqnarray}
and the commutation relation for the boson, but the statistics
is not important in our application below.  The Fock states are
defined by 
\begin{eqnarray}
|l\rangle=\hat{c}^\dag_l(0)|0\rangle
\end{eqnarray}
 with the vacuum state $|0\rangle$ satisfying $\hat{c}_l(0)|0\rangle=0$. 
The solution of the conventional Schr\"{o}dinger equation
\begin{eqnarray}
i\hbar\frac{\partial}{\partial t}\psi(t,\vec{x})=\hat{H}(t)\psi(t,\vec{x})
\end{eqnarray}
with the initial condition $\psi(0,\vec{x})=v_{n}(0,\vec{x})$
is given by 
\begin{eqnarray}
\psi_{n}(t,\vec{x})=\langle 0|\hat{\psi}(t,\vec{x})\hat{c}^{\dagger}_{n}(0)|0\rangle
\end{eqnarray}
by noting the operator equation $i\hbar\frac{\partial}{\partial t}\hat{\psi}(t,\vec{x})=\hat{H}(t)\hat{\psi}(t,\vec{x})$ and the equal-time commutation relation.

One now recognizes that this second quantized formulation contains the following gauge (or redundant) freedom
\begin{eqnarray}
\hat{c}_{n}(t)\rightarrow e^{-i\alpha_{n}(t)}\hat{c}_{n}(t),
\hspace{5mm} v_{n}(t)\rightarrow e^{i\alpha_{n}(t)}v_{n}(t)
\end{eqnarray}
where the phase freedom $\{\alpha_{n}(t)\}$ are arbitrary functions of time. Under this {\em hidden gauge transformation} in the second quantization, the field variable 
$\hat{\psi}(t,\vec{x})$ in (2) and thus the starting action $S$, the orthonormality of $\{v_{n}(t,\vec{x})\}$ and the (anti-)commutation relations of $\{\hat{c}_{n}(t)\}$ all remain invariant. But this transformation is not trivial. Under this gauge transformation, the solution of the Schr\"{o}dinger equation (7) is transformed as 
\begin{eqnarray}
\psi_{n}(t,\vec{x})=\langle 0|\hat{\psi}(t,\vec{x})\hat{c}^{\dagger}_{n}(0)|0\rangle\rightarrow e^{i\alpha_{n}(0)}\psi_{n}(t,\vec{x}).
\end{eqnarray}
This transformation by a time-independent phase factor corresponds to the {\em ray representation}, namely, behind the ray representation of the state vector in the Hilbert space we have an enormous hidden gauge freedom. In the analysis of geometric phases below, it is crucial that the combination 
\begin{eqnarray}
\psi_{n}^{\star}(0,\vec{x})\psi_{n}(t,\vec{x})
\end{eqnarray}
and thus its phase ${\rm arg}\{\psi_{n}^{\star}(0,\vec{x})\psi_{n}(t,\vec{x})\}$ is manifestly gauge invariant and thus physical. The hidden gauge symmetry is also crucial to define parallel transport and 
holonomy, as is shown in Section 4.

By inserting the expansion (2) into the action $S$, we obtain the effective Hamiltonian
\begin{eqnarray}
\hat{H}_{\text{eff}}(t)&=&\sum_{n,m}\int d^{3}x[ v^{\star}_{n}(t,\vec{x})\hat{H}(t)v_{m}(t,\vec{x})\nonumber\\
&&-v^{\star}_{n}(t,\vec{x})i\hbar\partial_{t}v_{m}(t,\vec{x})]\hat{c}^{\dagger}_{n}(t)\hat{c}_{m}(t).
\end{eqnarray}
 By noting the Heisenberg equation of motion
\begin{equation*}
i\hbar\frac{\partial}{\partial
t}\hat{c}_l(t)=[\hat{c}_l(t),\hat{H}_{\text{eff}}(t)],
\end{equation*}
it is confirmed that one can write
\begin{equation*}
\hat{c}_l(t)=U^{\dagger}(t)\hat{c}_l(0)U(t)
\end{equation*}
by introducing the ``Schr\"{o}dinger picture" effective Hamiltonian (by replacing $\hat{c}_{n}(t)\rightarrow \hat{c}_{n}(0)$ in $\hat{H}_{\text{eff}}(t)$)
\begin{eqnarray} 
\hat{\mathcal{H}}_{\text{eff}}(t)&\equiv&\sum_{n,m}\int d^{3}x[ v^{\star}_{n}(t,\vec{x})\hat{H}(t)v_{m}(t,\vec{x})\nonumber\\
&&-v^{\star}_{n}(t,\vec{x})i\hbar\partial_{t}v_{m}(t,\vec{x})]\hat{c}^{\dagger}_{n}(0)\hat{c}_{m}(0)
\end{eqnarray}
and the second-quantized formula of the evolution operator defined by
\begin{equation}\label{eq-evolution}
U(t)=\mathcal{T}\exp[-(i/\hbar)\int^t_0\hat{\mathcal{H}}_{\text{eff}}(t')
dt']
\end{equation}
where $\mathcal{T}$ represents a time ordered product. 
The Schr\"{o}dinger amplitude is then written as 
\begin{eqnarray}
\psi_{n}(t,\vec{x})&=&\langle 0|\hat{\psi}(t,\vec{x})\hat{c}^{\dagger}_{n}(0)|0\rangle\nonumber\\
&=&\sum_{m}v_{m}(t,\vec{x})\langle 0|\hat{c}_{m}(0)U(t)\hat{c}^{\dagger}_{n}(0)|0\rangle\nonumber\\
&=&\sum_{m}v_{m}(t,\vec{x})\langle m|U(t)|n\rangle
\end{eqnarray} 
and it is shown that the adiabatic approximation
corresponds to an approximate diagonalization of
$\hat{\mathcal{H}}_{\text{eff}}(t)$ \cite{2008Fujikawa}.

\section{Exactly solvable example}

In this section, we discuss an exactly solvable example associated with the motion of a spin inside the rotating magnetic field. This simple example \cite{2007Fujikawa,2008Fujikawa} is useful to illustrate a unified treatment of both of the adiabatic and non-adiabatic geometric phases sketched in Section 2.

We denote a rotating background magnetic field by
\begin{eqnarray}
 {\bm
B}(t)=B\big(\sin\theta \cos\varphi(t),
\sin\theta\sin\varphi(t),\cos\theta\big)
\end{eqnarray}
 and $\varphi(t) =\omega_0
t$ with a constant angular velocity $\omega_0$. The action for the spin
system in the second quantized formulation  is written as 
\begin{eqnarray}\label{eq-action1}
S&=&\int dt\left[\hat{\psi}^\dag(t)\Big(i\hbar\frac{\partial}{\partial
t}+{\bm B}\cdot {\bm \sigma}/2\Big)\hat{\psi}(t)\right],
\end{eqnarray}
with ${\bm \sigma}$ standing for the Pauli matrix and the field operator is expanded as
\begin{eqnarray}
\hat{\psi}(t,\vec{x})=\sum_{l=\pm}\hat{c}_l(t)w_l(t)
\end{eqnarray}
 with the anti-commutation relation,
$\big\{\hat{c}_l(t),\hat{c}_m^\dag(t)\big\}=\delta_{lm}$. The Fock states are defined by
$|l\rangle=\hat{c}^\dag_l(0)|0\rangle$ with the vacuum state
$|0\rangle$ satisfying $\hat{c}_l(0)|0\rangle=0$. 

For the above specific magnetic field with time-independent
$\theta$, the effective Hamiltonian for the isolated spin system is exactly diagonalized and, in this sense the spin system is exactly solvable if
one chooses the basis vectors as
\begin{eqnarray}
w_{+}(t)=\left(\begin{array}{c}
             e^{-i\varphi(t)}\cos\frac{\vartheta}{2}\\
            \sin\frac{\vartheta}{2}
            \end{array}\right),\
w_{-}(t)=\left(\begin{array}{c}
             e^{-i\varphi(t)}\sin\frac{\vartheta}{2}\\
            -\cos\frac{\vartheta}{2}
            \end{array}\right)
\end{eqnarray}
with $\vartheta=\theta-\theta_0$ and the constant parameter
$\theta_0$ defined by
\begin{equation}\label{eq-xi}
\tan\theta_0=\frac{\hbar\omega_0 \sin\theta}{B+\hbar\omega_0\cos\theta}.
\end{equation}
Then we have
\begin{eqnarray}\label{eq-weigen}
&&w_{\pm}^{\dagger}(t)\hat{H}w_{\pm}(t)
=\mp \frac{1}{2}B\cos\theta_0,\nonumber\\
&&w_{\pm}^{\dagger}(t)i\hbar\partial_{t}w_{\pm}(t)
=\frac{1}{2}\hbar\omega_0[1\pm\cos(\theta-\theta_0)],
\end{eqnarray}
with $\hat{H}=-{\bm B}(t)\cdot {\bm \sigma}/2$. In the operator
formulation of the second quantized theory, we obtain a
diagonalized effective Hamiltonian,
\begin{eqnarray}
\hat{H}_{\rm{eff}}(t)=\sum_{l=\pm} E_l \hat{c}_l^\dag(t)
\hat{c}_l(t)
\end{eqnarray}
where two time-independent effective energy eigenvalues are given by
\begin{eqnarray}\label{eq-Epm}
E_\pm&=&w_{\pm}^{\dagger}(t')\big(\hat{H}
-i\hbar\partial_{t'}\big)w_{\pm}(t')\nonumber\\
&=&
\mp\frac{1}{2}B\cos\theta_0-\frac{1}{2}\hbar\omega_0\big[1\pm\cos(\theta-\theta_0)\big].
\end{eqnarray}
By noting the Heisenberg equation of motion,
one can then write
\begin{equation}
\hat{c}_l(t)=U^{\dagger}(t)\hat{c}_l(0)U(t)
\end{equation}
by introducing the ``Schr\"{o}dinger picture" effective Hamiltonian $\hat{\mathcal{H}}_{\text{eff}}(t)\equiv\sum_lE_l\hat{c}_l^\dag(0)\hat{c}_l(0)$ and the second-quantized formula of the evolution operator defined by
$U(t)=\mathcal{T}\exp[-(i/\hbar)\int^t_0\hat{\mathcal{H}}_{\text{eff}}(t')
dt']$,
where $\mathcal{T}$ represents a time ordered product. 

For the Schr\"{o}dinger equation
$i\hbar\partial_{t}\psi_{\pm}(t)=\hat{H}\psi_{\pm}(t)$ with initial
condition $\psi_\pm(0)=w_\pm(0)$, its {\em exact} solution is given
in the second quantized notation \cite{2007Fujikawa, 2008Fujikawa}
\begin{eqnarray}\label{eq-exactamplitude}
\psi_{\pm}(t)&=&\langle
0|\hat{\psi}(t)\hat{c}^\dag_\pm(0)|0\rangle\nonumber\\
&=&\sum_lw_l(t)\langle 0|\hat{c}_l(0)U(t)\hat{c}^\dag_\pm(0)|0\rangle\nonumber\\
&=&w_{\pm}(t)\exp\left[-\frac{i}{\hbar}\int_{0}^{t}dt'
w_{\pm}^{\dagger}(t')\big(\hat{H}
-i\hbar\partial_{t'}\big)w_{\pm}(t')\right],\nonumber\\
\end{eqnarray}
where the exponent has been calculated in Eq. (\ref{eq-Epm}). Since
$w_{\pm}(T)=w_{\pm}(0)$ with the period $T=2\pi/\omega_0$, the
solution is {\em cyclic} \cite{1987Aharonov} (namely, periodic up to a phase freedom) and, as an exact solution, it is applicable to the
non-adiabatic case also. 

For an arbitrary time-dependent ${\bm
B}(t)$, any exact solution of the Schr\"{o}dinger equation can be
written in the last form of Eq.(\ref{eq-exactamplitude}) , if one
chooses basis vectors $w_{\pm}(t)$ suitably \cite{2007Fujikawa}. But
the periodicity $w_{\pm}(T)=w_{\pm}(0)$ is generally lost and thus
the solution becomes non-cyclic \cite{1988Samuel}.

At the adiabatic limit $|\hbar\omega_0/ B|\ll 1$, $\theta_0$ in
Eq. (\ref{eq-xi}) approaches zero so that the Schr\"{o}dinger 
amplitude approaches
\begin{eqnarray}\label{eq-exactamplitude2}
\psi_{\pm}(t)
&=&w_{\pm}(t)\exp\left[-\frac{i}{\hbar}[-\frac{1}{2}\hbar\omega_0\big(1\pm\cos(\theta-\theta_0)\big)t\right]\nonumber\\
&&\hspace{6mm}\times\exp\left[-\frac{i}{\hbar}[
\mp\frac{1}{2}B\cos\theta_0]t\right]\nonumber\\
&\rightarrow&w_{\pm}(t)\exp\left[\frac{i}{2}[\omega_0\big(1\pm\cos\theta \big)t\right]\nonumber\\
&&\hspace{6mm}\times\exp\left[-\frac{i}{\hbar}[
\mp\frac{1}{2}B]t\right]
\end{eqnarray}
where the first phase factor is usually called as the {\em geometric phase} and the second phase factor as the {\em dynamical phase}. 
The conventional Berry's phase  
$\pi(1\pm\cos\theta)$  \cite{1984Berry} or 
\begin{eqnarray}
\exp{[i\pi(1\pm\cos\theta)]}
\end{eqnarray}
is recovered after one cycle $t=T=2\pi/\omega_{0}$ of motion. This Berry's phase is known to have a topological meaning as the phase generated by a magnetic monopole located at the origin of the parameter space ${\bm B}$. Note that the dynamical phase in (\ref{eq-exactamplitude2}) vanishes at ${\bm B}=0$, namely, the level crossing appears in the conventional adiabatic approximation. This (potential) level crossing combined with exact adiabaticity is responsible for the topological property of Berry's phase. 

 On the other hand, at the non-adiabatic
limit $|\hbar\omega_0/ B|\gg 1$, $\theta_0$ approaches $\theta$ in Eq. (\ref{eq-xi}) so that the geometric phase in Eq. (\ref{eq-exactamplitude2}) vanishes. Namely, the adiabatic Berry's phase is smoothly connected to the trivial phase inside the exact
solution and thus the topology of Berry's phase is actually trivial\cite{2007Fujikawa, 2008Fujikawa}. In our unified formulation of adiabatic and non-adiabatic phases, we can analyze a transitional
region from the adiabatic limit to the non-adiabatic region in a reliable way, which was not possible in the past formulation.

One can assign a gauge invariant meaning to the geometric phase  under general
adiabatic or non-adiabatic evolution. To see this, let us recall
that the field variable
$\hat{\psi}(t,\vec{x})=\sum_{l=\pm}\hat{c}_l(t)w_l(t)$ in Eq.(\ref{eq-action1}) is invariant under the simultaneous replacements
\cite{2005Fujikawa}
\begin{equation}\label{eq-hiddenlocal}
\hat{c}_l(t)\rightarrow e^{-i\alpha_l(t)}\hat{c}_l(t),\ \
w_l(t)\rightarrow e^{i\alpha_l(t)}w_l(t),
\end{equation}
and thus the basic action Eq.(\ref{eq-action1}) is invariant under this
exact gauge symmetry. One then confirms that the exact Schr\"{o}dinger
amplitude $\psi_l(t)=\langle 0|\hat{\psi}(t)\hat{c}^{\dagger}_l(0)
|0\rangle$ in Eq.(\ref{eq-exactamplitude}) is transformed under
this gauge symmetry as $\psi_l(t)\rightarrow
\exp{[i\alpha_l(0)]}\psi_l(t)$ independently of $t$. The product
\begin{eqnarray}
\psi^{\dag}_l(0)\psi_l(t)
\end{eqnarray}
 is thus manifestly gauge invariant. Its
phase ${\rm arg}\{\psi^{\dag}_l(0)\psi_l(T)\}$ after subtracting the gauge invariant ``dynamical phase"
$\int_0^T dtw_l^{\dagger}(t)\hat{H}w_l(t)$ becomes
\begin{eqnarray}\label{eq-GP}
\beta_l ={\rm arg}\left\{
w^{\dag}_l(0)w_l(T)\exp\left[i\int_{0}^{T}dt
w_l^{\dagger}(t)i\partial_{t}w_l(t)\right]\right\},
\end{eqnarray}
which is also manifestly gauge invariant. This  $\beta_l$ agrees with the Berry's phase (26) in the adiabatic limit with $w_l(0)=w_l(T)$. This  $\beta_l$ is understood as the
holonomy of the {\em basis vector} associated with the exact hidden
local symmetry Eq.(\ref{eq-hiddenlocal}) for all geometric phases, either adiabatic or non-adiabatic, as is explained in the next section. This construction of holonomy is a generalization of Berry's phase to
the generic case of the non-vanishing dynamical phase, for which the Schr\"{o}dinger
amplitude does not satisfy the parallel transport condition analyzed by Simon\cite{simon}, but the
basis vector can satisfy the parallel transport condition with the
help of the gauge symmetry Eq.(\ref{eq-hiddenlocal})
\cite{2007Fujikawa}. For the non-cyclic case, one can still identify
Eq.(\ref{eq-GP}) as a gauge invariant non-cyclic geometric phase
\cite{1988Samuel}.

We here briefly compare the above unified formulation of geometric phases  to the
conventional formulation where the adiabatic phase is defined to be
invariant under the symmetry identical to the above hidden symmetry
(\ref{eq-hiddenlocal}) whereas the non-adiabatic phase is defined to
be invariant in the so-called {\em projective Hilbert space} with the
equivalence class \cite{1987Aharonov,1988Samuel}
\begin{eqnarray}
\{e^{i\alpha(t)}\psi(t)\}
\end{eqnarray}
where $\psi(t)$ stands for the conventional Schr\"{o}dinger amplitude. It is obvious that the above equivalence class or gauge symmetry $\psi(t)\rightarrow e^{i\alpha(t)}\psi(t)$ is not a symmetry of the Schr\"{o}dinger equation $i\hbar\partial_{t}\psi(t)=\hat{H}(t)\psi(t)$. As a consequence, the gauge
invariant non-adiabatic phase \cite{1987Aharonov,singh} on the basis of the projective Hilbert space 
\begin{eqnarray}\label{non-adiabatic}
\beta ={\rm
arg}\{\psi^{\dag}(0)\psi(T)\exp[i\int_{0}^{T}dt
\psi^{\dag}(t)i\partial_{t}\psi(t)]\}
\end{eqnarray}
  is
{\em non-local and non-linear} in the Schr\"{o}dinger amplitude $\psi(t)$, and thus consistency with the superposition principle is not obvious; in fact, it causes certain complications as was noted by Marzlin et al.
\cite{2004Marzlin}. 

In contrast, our $\beta_l$ in Eq. (\ref{eq-GP}),
which numerically agrees with Aharonov-Anandan's $\beta$ in (\ref{non-adiabatic}) when one
uses the exact solution Eq. (\ref{eq-exactamplitude}) in the definition of $\beta$, is bi-linear in the Schr\"{o}dinger amplitude and thus consistency with the superposition principle is manifest.

\section{Parallel transport and holonomy}

When the basis vector $v_{n}(\vec{x},t)$ is regarded as a complex number ($U(1)$ bundle), the parallel transport condition is defined by \cite{1992Anandan}
\begin{eqnarray}
\int d^{3}x v^{\dagger}_{n}(\vec{x},t)\partial_{t}v_{n}(\vec{x},t)=0.
\end{eqnarray}
This condition arises from the condition on the change of the phase under an infinitesimal time development  
\begin{eqnarray}
&&\int d^{3}x v^{\dagger}_{n}(\vec{x},t)v_{n}(\vec{x},t+\delta t)={\rm
real \ and \ positive}
\end{eqnarray}
and the normalization condition of the basis vector
\begin{eqnarray}
&&\int d^{3}x v^{\dagger}_{n}(\vec{x},t+\delta t)v_{n}(\vec{x},t
+\delta t)\nonumber\\
&&=\int d^{3}x v^{\dagger}_{n}(\vec{x},t)v_{n}(\vec{x},t)=1.
\end{eqnarray}
From the condition (33), the imaginary component of (32) vanishes, and from (34) the real component of (32) vanishes. 

Starting with a general $v_{n}(\vec{x},t)$, which does not satisfy the 
parallel transport condition, one may consider a specific gauge transformed 
basis vector 
\begin{eqnarray}
\bar{v}_{n}(\vec{x},t)=e^{i\alpha^{\prime}_{n}(t)}v_{n}(\vec{x},t).
\end{eqnarray}
Then the parallel transport condition
\begin{eqnarray}
\int d^{3}x \bar{v}^{\dagger}_{n}(\vec{x},t)\partial_{t}
\bar{v}_{n}(\vec{x},t)=0
\end{eqnarray}
fixes the gauge parameter $\alpha^{\prime}_{n}(t)$ as 
\begin{eqnarray}
\bar{v}_{n}(\vec{x},t)=
\exp[i\int_{0}^{t} dt^{\prime} d^{3}x^{\prime}
 v^{\dagger}_{n}(\vec{x}^{\prime},t^{\prime})i\partial_{t^{\prime}}
 v_{n}(\vec{x}^{\prime},t^{\prime})]v_{n}(\vec{x},t).
\end{eqnarray}
Under the hidden gauge transformation of the original basis vector 
$v_{n}(\vec{x},t)\rightarrow e^{i\alpha_{n}(t)}v_{n}(\vec{x},t)$, where the parameter $\alpha_{n}(t)$ is now arbitrary, we
have for $\bar{v}_{n}(\vec{x},t)$, which is non-linear in the original $v_{n}(\vec{x},t)$, 
\begin{eqnarray}
\bar{v}_{n}(\vec{x},t)\rightarrow e^{i\alpha_{n}(0)}
\bar{v}_{n}(\vec{x},t)
\end{eqnarray}
independently of $t$.

The {\em holonomy} is defined by the projection of the parallel transported $\bar{v}_{n}(\vec{x},t)$ after one-cycle $T$ (i.e., a periodic motion up to a phase) to the vector at $t=0$
\begin{eqnarray}
\bar{v}^{\star}_{n}(\vec{x},0)\bar{v}_{n}(\vec{x},T)
&=&v^{\star}_{n}(\vec{x},0)v_{n}(\vec{x},T)\nonumber\\
&\times&\exp[i\int_{0}^{T} dt^{\prime} d^{3}x^{\prime}
 v^{\dagger}_{n}(\vec{x}^{\prime},t^{\prime})i\partial_{t^{\prime}}
 v_{n}(\vec{x}^{\prime},t^{\prime})]\nonumber\\
 \end{eqnarray}
which is {\em manifestly gauge invariant}. This  holonomy defines the 
physical phase
\begin{eqnarray}
{\rm arg}\left\{v^{\star}_{n}(\vec{x},0)v_{n}(\vec{x},T)
e^{i\{\int_{0}^{T} dt^{\prime} d^{3}x^{\prime}
 v^{\dagger}_{n}(\vec{x}^{\prime},t^{\prime})i\partial_{t^{\prime}}
 v_{n}(\vec{x}^{\prime},t^{\prime})\}}\right\}\nonumber\\
\end{eqnarray}
which agrees with the geometric phase (29) when applied to the specific choice of the basis vector  $v_{n}(\vec{x},t)=w_{l}(t)$ in (18). For the periodic basis
$v_{n}(\vec{x},0)=v_{n}(\vec{x},T)$, the pre-factor $v^{\star}_{n}(\vec{x},0)v_{n}(\vec{x},T)$ is real and positive.

More generally, for {\em non-cyclic evolution} (i.e., $v_{n}(\vec{x},0)\neq v_{n}(\vec{x},t)$ for any finite $t$) one may define
\begin{eqnarray}
&&\int d^{3}x \bar{v}^{\star}_{n}(\vec{x},0)\bar{v}_{n}(\vec{x},t)
\nonumber\\
&=&\int d^{3}x v^{\star}_{n}(\vec{x},0)v_{n}(\vec{x},t)\nonumber\\
&\times&\exp[i\int_{0}^{t} dt^{\prime} d^{3}x^{\prime}
 v^{\dagger}_{n}(\vec{x}^{\prime},t^{\prime})i\partial_{t^{\prime}}
 v_{n}(\vec{x}^{\prime},t^{\prime})]
\end{eqnarray}
which is manifestly gauge invariant. By a suitable choice of the hidden local gauge, one can make the pre-factor $\int d^{3}x v^{\star}_{n}(\vec{x},0)v_{n}(\vec{x},t)$ real and positive. Then the phase
\begin{eqnarray}
{\rm arg}\left\{
\exp[i\int_{0}^{t} dt^{\prime} d^{3}x^{\prime}
 v^{\dagger}_{n}(\vec{x}^{\prime},t^{\prime})i\partial_{t^{\prime}}
 v_{n}(\vec{x}^{\prime},t^{\prime})]\right\}
\end{eqnarray}
 agrees with the non-cyclic and non-adiabatic geometric phase \cite{1988Samuel}. This phase for the non-cyclic evolution  is not called holonomy
in a precise sense. But it is convenient to include this quantity in 
the "holonomy" in a broad sense. In this case,
{\em all the known geometric phases} are understood as the holonomy
associated with the hidden local gauge symmetry.

\section{Robustness of geometric phases}

The geometric phase, in particular, the adiabatic Berry's phase exhibits 
topological properties in the ideal adiabatic limit. It is thus generally hoped that the geometric 
phase may provide a basis for a  device of quantum information processing which is robust against the change of external parameters. This idea is very interesting if it works \cite{1999Zanardi,2003Blais}.

This robustness issue has been recently examined on the basis of the second quantized formulation which gives a reliable treatment of geometric phases in the region away from  ideal adiabaticity \cite{2009Hu}. The basic idea of this analysis is to start with the exactly solvable example for a spin system in Section 3 and then incorporate the external perturbation or dissipation in the manner of Caldeira and Leggett\cite{1981Caldeira}. In this scheme one introduces an infinite set of 
harmonic oscillators to simulate the dissipation. The original formulation of Caldeira and Leggett uses the Euclidean path integral formulation, but for our purpose it is more convenient to use the second 
quantized formulation of the Caldeira-Leggett model\cite{1992Fujikawa}.  
We then evaluate the evolution operator of the spin system under the influence of the infinite number of external harmonic oscillators. The 
modification of the eigenvalues of the effective Hamiltonian is then interpreted as the effect of the dissipation on dynamical and geometric phases. 

The numerical analysis indicates that the deviation of the geometric phase from the ideal adiabatic value given by Berry
becomes sizable for $\hbar\omega_{0}/B\simeq 0.1$ and quite significant for
$\hbar\omega_{0}/B\simeq 0.5$. See (\ref{eq-exactamplitude2}). Any quantum device with the adiabatic speed of operation is not useful, and one may need to operate the device with a certain speed which is controlled by $\omega_{0}$. Our analysis may indicate that the robustness of Berry's phase against the non-adiabatic modification is rather fragile. 

A precise estimate of the effect of the dissipation on the geometric phase is not easy since one needs to know the rather detailed operational characteristics of the actual quantum device, but  the numerical analysis indicates that the effects are not completely negligible. This analysis of the effects of dissipation remain to be performed in more detail by taking into account the detailed specification of experiments.

As for the details of the analysis, the readers are referred to \cite{2009Hu}.

\section{Comparison with Aharonov-Bohm effect and chiral anomaly}

It is known that there are some similarities between the Aharonov-Bohm effect \cite{1959Aharonov}
and the adiabatic Berry's phase \cite{1984Berry}. It is however important to recognize one crucial difference between these two phases. The topology of the geometric Berry's phase is valid only in the ideal adiabatic limit and it is lost once one moves away from ideal adiabaticity, as was shown in Section 3. The topology of the adiabatic phase is self-generated by the singularity associated with the level crossing in the ideal adiabatic limit.

On the other hand, the topology of the Aharonov-Bohm effect is provided by the external boundary condition for the gauge field, and thus it is valid regardless of the speed of the motion of the electron. The Aharonov-Bohm phase is precise for the non-adiabatic  as well as adiabatic motion of the electron. This universal property characterizes the Aharanov-Bohm phase, and it makes the Aharanov-Bohm effect very useful in the variety of experiments. This difference between the Aharonov-Bohm effect and the Berry's phase is crucial when one considers practical applications. 

It is sometimes stated in the literature that the chiral quantum 
anomaly is regarded as an example of the geometric phase. 
Though there is some superficial similarity between these two 
notions, it is  shown that the differences between these two 
notions are more profound and fundamental \cite{2006Fujikawa}. It is shown that the 
geometric term associated with the Berry's phase, which
is topologically trivial for any finite time interval $T$ (i.e., away from ideal adiabaticity), 
corresponds to the so-called ``normal naive term'' in the analysis of chiral anomaly in field theory and has nothing to do with the 
anomaly-induced Wess-Zumino term. In the fundamental level, the difference between the two notions is stated as follows: The topology of 
gauge fields leads to level crossing in the fermionic sector
 in the case of chiral anomaly and the inevitable {\em failure} of the 
adiabatic approximation is essential in the analysis of chiral anomaly, whereas the (potential) level crossing in the matter sector leads to the 
topology of the Berry phase only when the adiabatic 
approximation precisely {\em holds}.  These two cannot be 
compatible with each other \cite{2006Fujikawa}.
\\

In the early literature on the geometric phase, the similarities 
between the geometric phase and other notions such as the quantum anomaly and Aharonov-Bohm effect were emphasized. That 
analogy was useful at the initial developing stage of the 
subject. But in view of the wide use of the 
terminology ``geometric phase'' in various fields in physics 
today, it is our opinion that similar phenomena should be clearly distinguished from the same phenomena. What we are suggesting is to call
chiral anomaly as chiral anomaly, and Aharovov-Bohm phase as Aharonov-Bohm phase, etc., since those terminologies convey  very clear 
 and well-defined physical contents. Even in this sharp 
definition of terminology, one should still be able to clearly identify the geometric (or Berry) phase and its physical characteristics, 
which cannot be described by other notions. 

\section{Conclusion}

We have shown that the second quantization induces a hidden local gauge symmetry and that this hidden symmetry defines the parallel transport and holonomy associated with geometric phases.
This approach allows a unified treatment of all the known geometric phases, either adiabatic or non-adiabatic, and thus one can analyze the 
transitional region from adiabatic to non-adiabatic phases in a reliable way.  One then recognizes that the topology of the adiabatic Berry's phase is actually trivial, and  the robustness of the geometric phase against the non-adiabatic modification is rather fragile contrary to a naive expectation.
We have also emphasized the basic difference between Berry's phase and 
Aharonov-Bohm phase. A formal similarity between the geometric phase and 
chiral anomaly was shown to be illusionary.

Nevertheless, we should emphasize that the existence of an extra phase 
in the adiabatic approximation is interesting. Moreover, it has been shown by Berry that  the extra phase, which is commonly called geometric phase, has observable physical effects. The full physical implications of geometric phases {\em per se} remain to be further exploited.   

We also believe that a sharp definition of the scientific term 
``geometric phase'' is 
 important for those experts working on the geometric phase 
itself, since then the wider audience can easily 
identify the phenomena, which are intrinsic to the geometric 
phase and  cannot be described by other notions, and  
consequently they will appreciate more the usefulness of the 
geometric phase.

\end{document}